\documentclass[preprint,showpacs,preprintnumbers,amsmath,amssymb]{revtex4}

\usepackage{graphicx}
\usepackage{dcolumn}
\usepackage{bm}

\begin{document}
\title{Tunneling conductance of a two-dimensional electron gas with Rashba
spin-orbit coupling}

\author{B. Srisongmuang and P. Pairor}
 \email{pairor@sut.ac.th}
\affiliation{School of Physics, Institute of Science,
        Suranaree University of Technology, Thailand}
\author{M. Berciu}
\affiliation{Department of Physics and Astronomy, University of
        British Columbia, British Columbia, Canada}
\date{\today}
\begin{abstract}
We theoretically studied the in-plane tunneling spectroscopy of the
hybrid structure composed of a metal and two-dimensional electron
gas with Rashba spin-orbit coupling. We found that the energy
spacing between two distinct features in the conductance spectrum
can be used to measure the Rashba energy.  We also considered the
effect that varying the probability of spin-conserving and spin-flip
scattering at the interface has on the overall conductance.
Surprisingly, an increase in interface scattering probability can
actually result in increased conductance under certain conditions.
Particularly, in the tunneling regime, an increase in spin-flip
scattering probability enhances the conductance. It is also found
that the interfacial scattering greatly affects the spin
polarization of the conductance in metal, but hardly affects that in
the Rashba system.
\end{abstract}

\pacs{73.40.Ns, 73.40.Gk, 73.23.-b, 72.25.Dc, 72.25.Mk}

\maketitle

\section{Introduction}
Structural inversion asymmetry of the confining electrostatic
potential results in an intrinsic spin-orbit coupling of electrons
in a two-dimensional (2D) electron gas (EG), which can be described
by the Rashba Hamiltonian~\cite{rashba,rashba2,rashba3}:
\begin{equation}
\mathcal{H} =
\frac{{\vec{p}}^2}{2m^*}-\lambda\hat{j}\cdot(\vec{p}\times\vec{\sigma})
\end{equation}
where $\vec{p}$ is 2D momentum, $m^*$ is the electron effective
mass, $\hat{j}$ is the direction perpendicular to the plane of
motion, $\lambda$ is the spin-orbit coupling parameter, which can be
tuned by applying an external gate voltage perpendicular to the 2D
plane, and the components of $\vec{\sigma}$ are the Pauli spin
matrices. The spin-orbit interaction lifts the spin degeneracy and
causes the original parabolic energy spectrum to split into two
branches: $E_{\vec{k},\pm} = \frac{\hbar^2k^2}{2m^*}\pm\hbar\lambda
k$, where $k$ is the magnitude of the wave vector. The density of
states of this system is the same as that of the 2D free electron
gas for all energies higher than the crossing point of the two
branches. However, at the bottom of the band, the density of states
has $E^{-\frac{1}{2}}$ van Hove singularity because the minus branch
has an annular minimum for $k=k_0\equiv m^*\lambda/\hbar$ instead of
a single-point minimum as in the free electron gas. These properties
lead to interesting phenomena, like the spin hall effect (see e.g.
Refs.~\cite{engel} for review), and to applications in spintronics
(see e.g. Refs.~\cite{zutic} for review).

The Rashba effect has been seen in many systems like surface alloys
and semiconductor heterostructures. Several techniques have been
used to study the spin-split states in these systems. For instance,
angle-resolved photoemission
spectroscopy~\cite{lashell,rein,henk,cer,popovic} and scanning
tunneling microscopy~\cite{ast} are used in surface alloys. The
former technique is utilized mainly to obtain the energy dispersion
and the Fermi surface map, from which the effective mass, the
magnitude of the band splitting, and hence the Rashba spin-orbit
coupling energy, $E_\lambda\equiv \hbar^2k_0^2/(2m^*)$, can be
extracted~\cite{lashell,rein,henk,cer,popovic}. In the latter
technique, the electric current is driven through a sharp tip
perpendicular to the 2D plane and the differential conductance
($dI/dV$) spectrum can be obtained. One can deduce the Rashba energy
by fitting the $dI/dV$ spectrum to the local density of states of
the 2DEG~\cite{ast}. In both cases, to obtain information about the
Rashba spin-orbit coupling, extensive data fitting is needed.

In semiconductor heterostructures, the Rashba energy is measured
using the Shubnikov-de Haas oscillations~\cite{nitta,engels}. The
existence of the spin splitting at the Fermi energy leads to beating
in the oscillations and the Rashba energy can be deduced from the
position of the beating node. However, because this technique is
done in the presence of magnetic field, it tends to provide an
overestimate of the Rashba energy because it includes the effect of
the Zeeman spin splitting~\cite{lommer}.

In this article, we propose a way to measure the spin-splitting
energy more directly from the experimental data: the in-plane
tunneling spectroscopy. In this technique, the Rashba energy can be
determined by the energy difference between two features in the
conductance spectrum. The required condition for the measurement is
that the energy resolution of the tunneling spectra is at least of
the order of the Rashba energy itself. This condition can be easily
achieved in modern tunneling measurements~\cite{wolf}.

An intriguing property of 2DEG with Rashba spin-orbit interaction is
spin-dependent transport. Many theoretical investigations have shown
that both electric and spin transport in hybrid structures between
the Rashba system and various materials, like
metals~\cite{liu,lee,mari}, ferromagnets~\cite{mari,hu2,hu,jiang},
and superconductors~\cite{tanaka}, are affected by the strength of
the spin-orbit coupling~\cite{liu,lee,mari,hu2,hu,jiang,tanaka}, the
inequality of the effective masses~\cite{liu,lee,hu,jiang}, and the
transparency of the interface~\cite{hu2,hu,tanaka}. However, in
these previous studies, only spin-conserving interfacial scattering
was considered. In the presence of interfacial spin-flip scattering,
the equations describing the spin-up and spin-down spin states are
coupled and one expects interesting consequences of this. For
instance, in the study of the tunneling conductance spectrum of a
semiconductor/superconductor junction~\cite{zutic2}, the
non-spin-flip scattering, when present alone, is found to suppress
the Andreev reflection process and hence the subgap conductance as
expected. However, when the spin-flip potential scattering is also
present at the interface, their combined effect surprisingly
enhances the subgap conductance~\cite{zutic2}.

Here, we consider how the scattering potential barrier affects both
the conductance spectrum and the spin polarization of the
conductance of a junction consisting of a metal and a Rashba system.
As in previous work by Zutic and Das Sarma~\cite{zutic2}, we find
that the conductance spectrum, which is usually suppressed in the
presence of the interfacial scattering, can be enhanced by the
combined effect of both types of scattering. We also find that the
spin polarizations of conductance of the metal and the Rashba system
are not equal. The spin polarization in the latter depends weakly on
interfacial scattering, while that in the former is greatly
affected. This suggests that a spin imbalance in the Rashba system
is robust against variation in the quality of the junction
interface.

In the next section, we describe the theoretical method and
assumptions. In Section~\ref{result}, we provide the results and
discussion. The conclusion is given in the last section.

\section{Method of calculation and assumptions} \label{method}
\begin{figure}
\begin{center}
\includegraphics[scale=0.4]{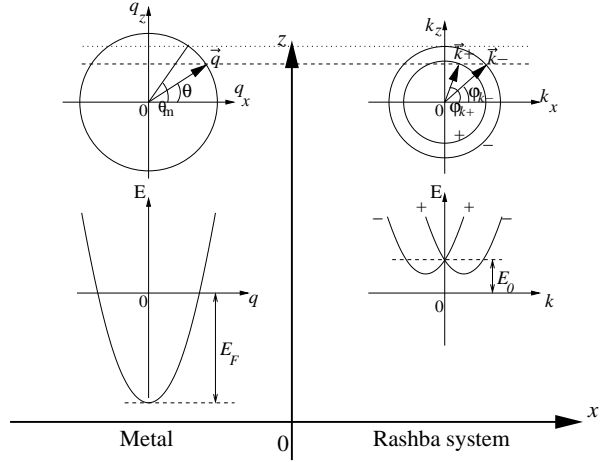}
\end{center}
\caption{\label{junction}The top sketches are the energy contours of
the electron in the metal (left) and the Rashba system (right). The
angles $\theta$ and $\varphi$ are defined as those between the $x$
axis and the momenta of electrons in the metal and the Rashba system
respectively. The dashed line that crosses both sides shows the
momentum states with the same $k_z$. The dotted line is the line of
the maximum value of $k_z$, which defines the maximum incident angle
$\theta_{m}$. The lower sketches are the corresponding energy
spectra ($E$ {\it vs} the magnitude of momentum). $E_F$ and $E_0$
are the metal Fermi energy and the off-set energy of the Rashba
system respectively.}
\end{figure}
We represent our junction by an infinite 2D system which lies on
$xz$ plane, where the metal and the Rashba system occupy the $x<0$
and $x>0$ region respectively. The two regions are separated by a
flat interface at $x=0$. The interfacial scattering is modeled by a
Dirac delta function potential~\cite{btk}. We consider ballistic
transport in our junction. In the one-band effective-mass
approximation, we describe our system by the following Hamiltonian:
\begin{equation}
\mathcal{H} = \left(\hat{p}\frac{1}{2m(x)}\hat{p} +
V(x,z)\right)\mathcal{I} + \mathcal{H}_R(x).
\end{equation}
Each term is the $2\times2$ matrix acting on spinor states.
$\hat{p}=-i\hbar\left(\hat{x}\frac{\partial}{\partial x} +
\hat{z}\frac{\partial}{\partial z}\right)$. The effective mass
$m(x)$ is position-dependent, i. e., $[m(x)]^{-1} =
m^{-1}\Theta(-x)+(m^*)^{-1}\Theta(x)$, where $m$ and $m^*$ are
effective electron masses in the metal and the Rashba system
respectively, and $\Theta(x)$ is the Heaviside step function.
$V(x,z)$ is also a position-dependent function and is modeled by the
expression
\begin{equation}
V(x,z) = H\delta(x) + E_0\Theta(x) - E_F\Theta(-x)
\end{equation}
where $H$ represents the scattering potential at the interface,
$E_0$ is the energy difference between the Fermi level and the
bottom of the plus branch (see FIG.~\ref{junction}), and
$E_F=\hbar^2q_F^2/(2m)$ is the Fermi energy of the metal. We assume
that $E_F$ is much larger than $E_0$. The diagonal elements of $H$,
$H_{\uparrow\uparrow}$ and $H_{\downarrow\downarrow}$ correspond to
the non-spin-flip scattering potential characterizing the quality of
the junction, while $H_{\uparrow\downarrow}=H_{\downarrow\uparrow}$
describe spin-flip scattering~\cite{zutic2}. The Rashba Hamiltonian
is written as~\cite{zulicke}
\begin{equation}
\mathcal{H}_R(x) = \frac{\hat{j}}{2} \cdot
\left[\lambda(x)\left(\vec{p}\times\vec{\sigma}\right) +
\left(\vec{p}\times\vec{\sigma}\right)\lambda(x)\right]
\end{equation}
where $\lambda(x)=\lambda\Theta(x)$.

From the Hamiltonian, one can obtain the eigenstates and eigenenergy
for the electrons in each region as follows. In the $x<0$ region,
the energy spectrum is
\begin{equation}
E(q)=\frac{\hbar^2q^2}{2m}-E_F
\end{equation}
where $q =\sqrt{q_x^2+q_z^2}$ is the magnitude of the 2D momentum of
the electrons. In the $x>0$ region, the eigenenergy is obtained as
\begin{equation}
E^{\pm}(k) = \frac{\hbar^2k^2}{2m^*} \pm \frac{\hbar^2k_0^2}{2m^*}
+ E_0
\end{equation}
where $k=\sqrt{k_x^2+k_z^2}$ is the magnitude of the 2D momentum and
$k_0=m^*\lambda/\hbar$. FIG.~\ref{junction} shows the energy spectra
and energy contours of the excitations in both sides of the
junction.

The wave function of the electrons in metal is written as a linear
combination of incoming momentum state with equal spin components
and a reflected state of the same energy and $k_z$:
\begin{equation}
\Psi_M(x,z)=\left(\frac{1}{\sqrt{2}} \left[\begin{array}{c}
1\\
1\end{array}\right]e^{iq_xx} + \left[\begin{array}{c}
b_{\uparrow}\\
b_{\downarrow}\end{array}\right]e^{-iq_xx}\right)e^{ik_zz}
\end{equation}
where the $b_{\uparrow}, b_{\downarrow}$ are the amplitudes for
reflection of spin-up and spin-down electrons respectively.
$q_x=q\cos\theta$ and $k_z=q\sin\theta$, where $\theta$ is the angle
between $\vec{q}$ and the $x$ axis. The magnitude of the momentum,
$q$, depends on energy as
\begin{equation}
q=\sqrt{\frac{2m}{\hbar^2}(-E+E_F)}
\end{equation}

Similarly, in the Rashba system, the wave function is obtained as a
linear combination of two outgoing eigenstates of the same energy
and $k_z$:
\begin{equation}~\label{2deg}
\Psi_{RS}(x,z)=\left(c_+\left[\begin{array}{c}
\cos\frac{\varphi_{k^+}}{2}\\
\pm\sin\frac{\varphi_{k^+}}{2}\end{array} \right]e^{\mp
ik_x^+x}\right.
+\left. c_- \left[\begin{array}{c}\sin\frac{\varphi_{k^-}}{2}\\
\cos\frac{\varphi_{k^-}}{2}\end{array}
\right]e^{ik_x^-x}\right)e^{ik_zz}
\end{equation}
where $\varphi_{k^\pm}$ are the angles between $\vec{k}^\pm$ and the
$x$ axis. For $E>E_0$, $c_+$ and $c_-$ are the transmission
amplitudes of electrons to plus and minus branch respectively. When
$E<E_0$, $c_+$ and $c_-$ refer to the transmission amplitudes of
electrons to states with smaller and larger $k$ of the minus branch
respectively. The upper and lower signs in the first term of Eq.
(\ref{2deg}) are for $E\leq E_0$ and $E>E_0$ respectively.
$k_x^{\pm}=k^{\pm}\cos\varphi_{k^\pm}$ and
$k_z=k^\pm\sin\varphi_{k^\pm}$, where the magnitudes of the momenta,
$k^\pm$, depend on energy as
\begin{eqnarray}
k^-&=&k_0+\sqrt{k_0^2+\frac{2m^*}{\hbar^2}(E-E_0)}\\
\label{kplus}
k^+&=&\pm\left(k_0-\sqrt{k_0^2+\frac{2m^*}{\hbar^2}(E-E_0)}\right)
\end{eqnarray}
Again, in Eq. (\ref{kplus}) the upper and lower signs are for $E\leq
E_0$ and $E>E_0$ respectively. The relationship between the angles
$\varphi_{k^\pm}$ and $\theta$ is
\begin{equation}
k^\pm\sin\varphi_{k^\pm}=q\sin\theta.
\end{equation}

We can obtain the probability amplitudes
$b_\uparrow,b_\downarrow,c_+$ and $c_-$ from the following matching
conditions that ensure probability conservation~\cite{zulicke}.
\begin{eqnarray}
\Psi_M(x=0,z)&=&\Psi_{RS}(x=0,z)\equiv\Psi_0,\\
\frac{m}{m^*}\left.\frac{\partial\Psi_{RS}}{\partial
x}\right|_{x=0}\left.-\frac{\partial\Psi_M}{\partial
x}\right|_{x=0}&=&\left(2q_F\mathcal{Z}-i\frac{m}{m^*}k_0\sigma_z\right)
\Psi_0,
\end{eqnarray}
where $\mathcal{Z}=mH/(\hbar^2q_F)$. The diagonal elements of
$\mathcal{Z}$ will henceforth be referred to as $Z_u\equiv
Z_{\uparrow\uparrow}$ and $Z_d\equiv Z_{\downarrow\downarrow}$,
while the off-diagonal element will be denoted by
$Z_F=Z_{\uparrow\downarrow}=Z_{\downarrow\uparrow}$. In what follows
the spin flip term $Z_F$ will be responsible for the enhancement of
a feature at the branch-crossing point in the conductance spectrum.

The particle current density along the $x$ direction is obtained
from
\begin{equation}
j^p_x =\frac{1}{2}\left[\Psi^\dagger(x)\hat{v}_x\Psi(x) +
\left(\hat{v}_x\Psi(x)\right)^\dagger\Psi(x)\right],
\end{equation}
where $\Psi(x)$ is the spinor wave function, and $\hat{v}_x =
d\hat{x}/dt = i\left[\mathcal{H}(x),\hat{x}\right]/\hbar$. From the
current density, the reflection and transmission probabilities can
be obtained:
\begin{eqnarray}
R_\uparrow &=& \left|b_\uparrow\right|^2\\
R_\downarrow &=& \left|b_\downarrow\right|^2\\
T_+ &=& \frac{m}{m^*}\left|c_+\right|^2\left(\frac{\mp
k_x^+ + k_0\cos\varphi_{k_x^+}}{q_x}\right)\\
T_- &=& \frac{m}{m^*}\left|c_-\right|^2\left( \frac{k_x^-
-k_0\cos\varphi_{k_x^-}}{q_x}\right)
\end{eqnarray}
where $R_{\uparrow},R_{\downarrow}$ are the reflection probabilities
to spin-up and spin-down states, and $T_+,T_-$ are the corresponding
transmission probabilities. Also, the upper and lower signs in $T_+$
are for $E\leq E_0$ and $E>E_0$ respectively. As mentioned earlier,
the matching conditions ensure that
$R_\uparrow+R_\downarrow+T_++T_-=1$.

Since the current is independent of $x$, we consider the current
density in the metal for simplicity. It can be written as a function
of applied voltage $V$ as follows.
\begin{equation}
j^e_x(eV)=\sum_{q_x>0,q_z}ev_x\left(1-R_\uparrow-
R_\downarrow\right)\left[f(E_q-eV)-f(E_q)\right]
\end{equation}
where $e$ is the electron charge, $v_x$ is the $x$ component of the
electron group velocity, and $f(E)$ is Fermi distribution function.

By changing the integration variable and setting temperature to zero
for simplicity, one can obtain the expression for the electric
current as
\begin{equation}
j^e_x(eV) = \frac{e}{h}\frac{\mathcal{L}^2q_F}{2\pi}\int_0^{eV}
dE\int_{-\theta_{m}}^{\theta_{m}}d\theta \,\cos\theta
 \sqrt{1-\frac{E}{E_F}} \left(1-R_\uparrow-
R_\downarrow\right)
\end{equation}
where $\mathcal{L}^2$ is the area of the metal and $\theta_m$ is the
maximum angle of the incident electrons from the metal (see
FIG.~\ref{junction}): $\theta_m=\sin^{-1}(k^-(E)/q(E))$. Thus, the
differential conductance $G(V) \equiv dj^e_x/dV$ at zero temperature
is
\begin{equation}
G(V)= \frac{e^2}{h}\frac{\mathcal{L}^2q_F}{2\pi}
 \int_{-\theta_{m}}^{\theta_{m}}d\theta \,\cos\theta
 \sqrt{1-\frac{eV}{E_F}} \left(1-R_\uparrow -
R_\downarrow\right)
\end{equation}
The finite temperature will smear the features in the conductance
spectrum but will not change their positions (assuming that the
strength of the Rashba spin-orbit coupling does not depend on
temperature).

The spin polarization of the conductance $\mathcal{P}$ is defined as
\begin{equation}
\mathcal{P}(E)=\frac{\sum_{q_x>0,q_z}\left(n_\uparrow v_\uparrow -
n_\downarrow v_\downarrow\right)} {\sum_{q_x>0,q_z}\left(n_\uparrow
v_\uparrow + n_\downarrow v_\downarrow \right)},
\end{equation}
where $n_\sigma v_\sigma$ is the number of electrons with spin
$\sigma$ that cross a plane of interest per unit time. In terms of
the reflection probabilites this spin polarization in the metal can
be written as
\begin{equation}
\mathcal{P}_M(E)=\frac{\sum_{q_x>0,q_z}\left(-R_\uparrow +
R_\downarrow\right)} {\sum_{q_x>0,q_z}\left(1-R_\uparrow -
R_\downarrow\right)}
\end{equation}
and the spin polarization in the Rashba system in terms of the
transmission probabilities is written as
\begin{equation}
\mathcal{P}_{RS}(E)=\frac{\sum_{q_x>0,q_z}\left(T_+\cos\varphi_{k^+}
- T_-\cos\varphi_{k^-}\right)} {\sum_{q_x>0,q_z}\left(T_+ +
T_-\right)}
\end{equation}
As expected, the spin polarization measures the difference in number
of the carriers with spin-up and spin-down on each side.
\begin{figure}\label{comp}
\begin{center}
\includegraphics[scale=0.37]{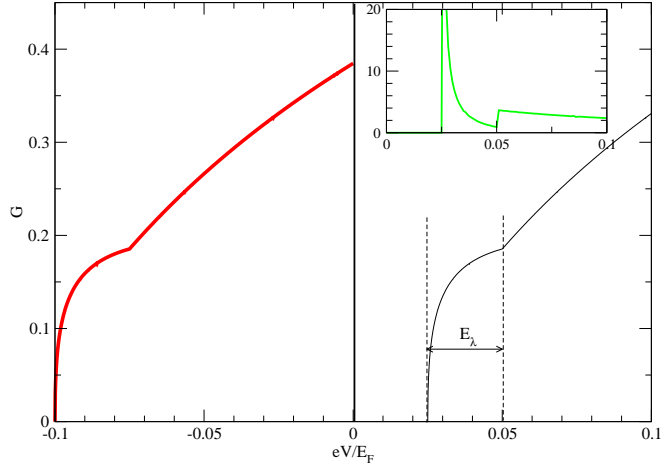}
\end{center}
\caption{\label{cond-comp} On the left side is the spectrum in the
case where the energy band of the Rashba system is occupied
($E_0=-0.075E_F$) and on the right is that where the band is
unoccupied ($E_0=0.05E_F$). $Z$ and $Z_F$ are set equal to zero.
$m/m^*=10$ and $k_0=0.05q_F$. The derivative of the conductance
spectrum on the right ($dG/dV$) is shown in the inset.}
\end{figure}

\section{Results and discussion}\label{result}
In this section, we discuss the effect of the interfacial scattering
on the differential conductance spectra and the spin polarization of
conductance on each side of the junction. In all plots, we set
$m/m^*=10$ and $k_0=0.05q_F$, which corresponds to typical
experimental values in metal/Rashba system
junctions~\cite{zutic2,koga}. In FIG.~\ref{cond-comp}, two
conductance plots for two values of $E_0$ are shown. Positives
values of $E_0$ means the energy spectrum of the Rashba system is
unoccupied and the positive $eV$ across the junction will cause the
current to flow from the metal to the Rashba system.

In the case shown in FIG.~\ref{comp} where the energy spectrum is
occupied ($E_0=-0.075E_F$), the results are identical in shape to
those in the unoccupied case ($E_0=+0.05E_F$), but the applied
voltage $eV$ across the junction has to be negative. There are two
main features at the voltage corresponding to the bottom and the
branch-crossing of the energy band. The distance between them
depends on $E_\lambda$, which is the quantity of interest. The value
of $E_0$ is not important, i. e., changing $E_0$ causes a rigid
shift in energy, and will henceforth be set equal to zero.

We do not consider the spin filtering interface. That is, we set the
non-spin-flip scattering strength $Z_u=Z_d=Z$. It is well-known that
the difference in $Z_u$ and $Z_d$ will cause a spin-filtering
effect. That is, a higher $Z_u$ will make the transport of the
spin-up electrons less favorable and vice versa. This effect cannot
be seen in the conductance spectrum and will not be considered in
the spin polarization.
\begin{figure}
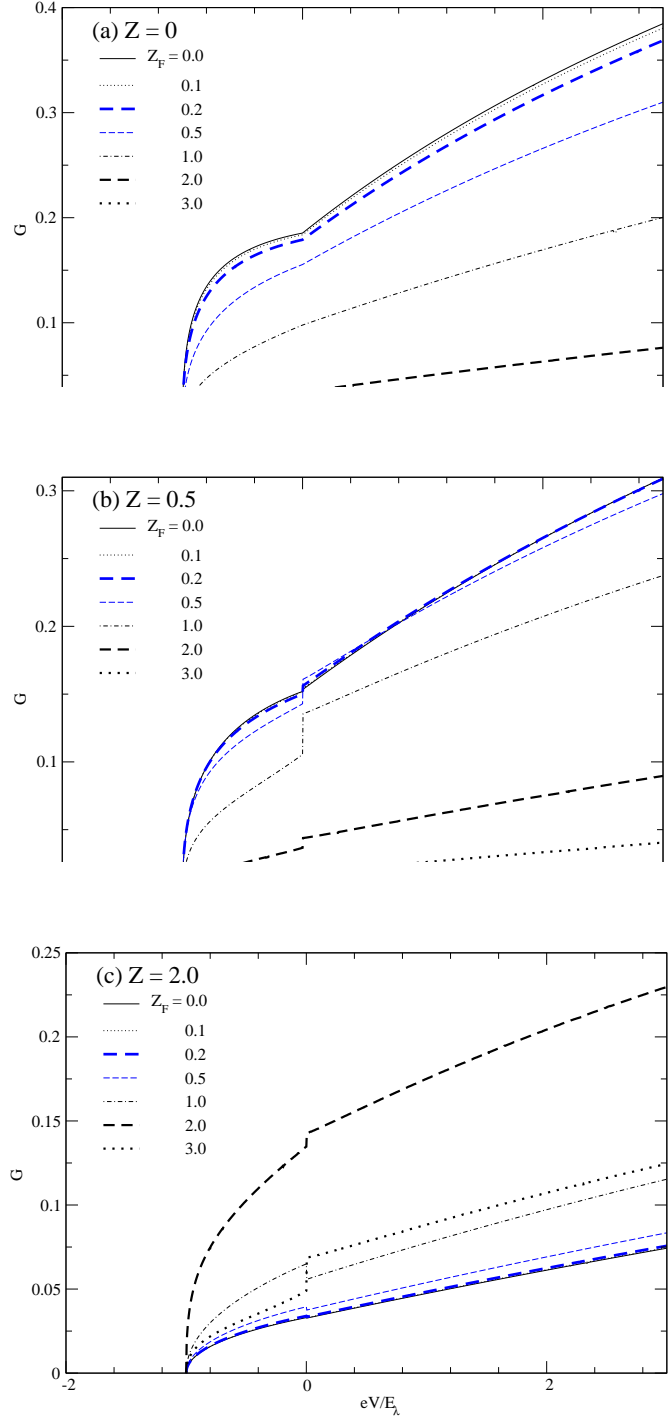

\begin{center}
\includegraphics[scale=0.37]{cond-z-0-fig3a.eps}\\
\includegraphics[scale=0.37]{cond-z-05-fig3b.eps}\\
\includegraphics[scale=0.37]{cond-z-20-fig3c.eps}
\end{center}
\caption{\label{cond} Differential conductance spectra $G$ for
different $Z_F$ in the case where (a) $Z=0$, (b) $Z=0.5$, and (c)
$Z=2.0$.}
\end{figure}

\subsection{Differential conductance spectra}
In all conductance plots, the conductance is in units of
$e^2\mathcal{L}^2q_F/(2\pi h)$. The conductance spectra $G$ with
different $Z_F$ in different limits of $Z$ are shown in
FIG.~\ref{cond}. Junctions with metallic contacts are characterized
by $Z\ll1$, whereas those in the tunneling limit are characterized
by $Z\ge1$. In general, the conductance is zero until the applied
voltage reaches $eV=-E_\lambda$, which is the bottom of the band of
the Rashba system. The conductance increases suddenly with large
initial slope that decreases steadily until a second feature: the
kink occurring at $eV=0$, which is the crossing point of the two
branches of the band. After this point, the conductance increases
approximately linearly. In the presence of $Z_F$, there occurs a
discontinuity in the conductance at $eV=0$. The height of the jump
depends on both $Z$ and $Z_F$. The energy difference between the
onset and the discontinuity in slope of the conductance spectrum can
be used to measure the magnitude of the Rashba energy $E_\lambda$.

In addition to the influence on the discontinuity at $eV=0$, the
interfacial scattering affects the overall conductance spectrum as
well. For metallic contacts, the spin-flip scattering suppresses the
conductance as expected. However, in the intermediate and the
tunneling limits, the results are rather surprising. As can be seen
in FIG.~\ref{cond}(b) when $Z=0.5$, the increase in $Z_F$ from zero
to a small value (less than 0.5) does not affect the conductance
much. Only when $Z_F$ is increased beyond 0.5, does the conductance
get suppressed. When $Z$ is high, e. g. $Z=2.0$ as in
FIG.~\ref{cond}(c), the conductance spectrum can be enhanced by the
increase in $Z_F$ up to a value $Z_F^*$, after which the spectrum
becomes suppressed. $Z_F^*$ is found to depend strongly on $Z$.
\begin{figure}
\begin{center}
\includegraphics[scale=0.33]{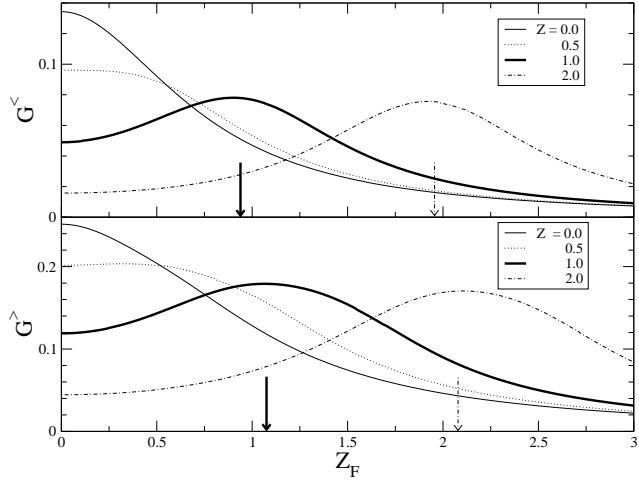}
\end{center}
\caption{\label{const-z} Differential conductance $G(eV)$ plotted as
a function of the spin-flip barrier height $Z_F$ at a constant
energy $eV$ slightly below [upper panel, denoted by $G^<(Z_F)$] and
slightly above [lower panel, denoted by $G^>(Z_F)$)] the energy
corresponding to the crossing of the Rashba-split bands. The arrows
indicate the values of $Z_F^*$, where the maximum differential
conductances $G^>$ and $G^<$ occur, for $Z=1.0$ (thick arrows) and
2.0 (dashed-dotted arrows).}
\end{figure}

One can see the effect on the conductance spectrum of spin-flip
scattering more clearly by considering plots of the conductance $G$
as a function of $Z_F$ for energies just below and just above $0$.
In FIG.~\ref{const-z}, $G^<\equiv G(-\delta)$ and $G^>\equiv
G(+\delta)$, where $\delta/E_\lambda=0.8$, are plotted as a function
of $Z_F$ for different values of $Z$. For small $Z$, both $G^>$ and
$G^<$ decrease with $Z_F$ as expected. However, this trend starts to
change when $Z$ is higher than 0.5. That is, both $G^>$ and $G^<$
increase with $Z_F$ and reach a maximum value at $Z_F^*$ (as
indicated by arrows in FIG.~\ref{const-z}), after which they
decreases with $Z_F$. Notice that $Z_F^*$ is a little smaller for
$G^<$ than for $G^>$ and is approximately equal to $Z$. It should be
noted that a similar dependence of both $G^>$ and $G^<$ on $Z$ can
also be seen, if one plots $G^>$ and $G^<$ as a function of $Z$
instead.
\begin{figure}
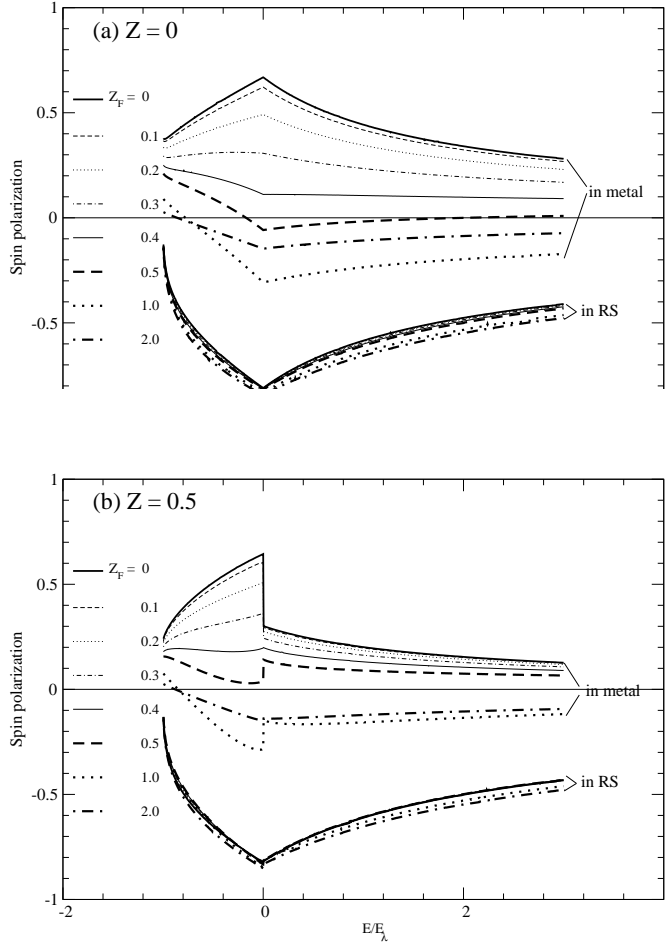

\begin{center}
\includegraphics[scale=0.37]{spolar-z-0-fig5a.eps}\\
\includegraphics[scale=0.37]{spolar-z-05-fig5b.eps}
\end{center}
\caption{\label{spolar-z0}The plots of the spin polarization of
conductance in metal and Rashba system (RS) as a function of energy
when $Z$ is (a) 0 and (b) 0.5.}
\end{figure}

\subsection{Spin polarization of conductance}
\begin{figure}
\begin{center}
\includegraphics[scale=0.38]{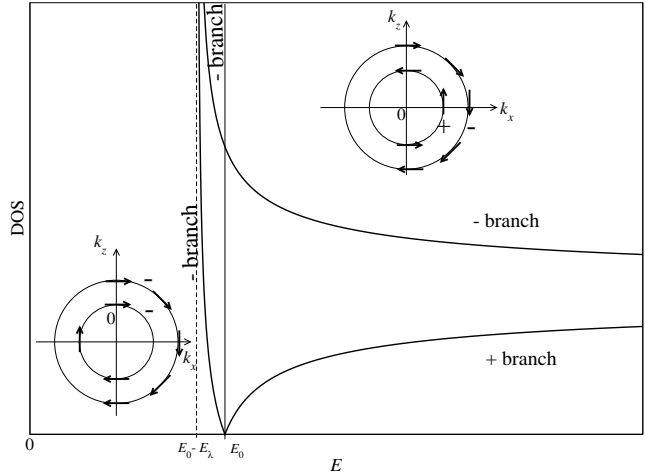}
\end{center}
\caption{\label{dos}Density of states of each branch of the 2DEG
with the Rashba spin-orbit coupling. The contour plots on the left
and on the right are those in the case where $E<E_0$ and $E>E_0$
respectively. When $E>E_0$, the outer contour is that of - branch
and the inner one is that of + branch. When $E<E_0$, both energy
contours belong to the - branch. The arrows represent the spin
direction of the states with positive $v_x$.}
\end{figure}
The plots of the spin polarizations in both metal and Rashba system
as a function of energy are shown in FIG.~\ref{spolar-z0}. The spin
polarizations of the two sides are very different. In Rashba system,
it is always negative, whereas in the metal it is positive when the
spin-flip scattering is not strong. This may be understood by
considering the density of states of the Rashba system.

The density of states of the minus branch is larger than that of the
plus branch. As we can see from FIG.~\ref{dos}, because the spins of
the transmitted states of the minus branch are mostly pointing down,
it is not surprising that the spin polarization in the Rashba system
is negative. As for the metal side, because more spin-down states
are transmitted into the Rashba system, the spin polarization is
positive.

The interfacial scattering does not affect the spin polarization in
the Rashba system as much as in the metal. The increase in either
$Z$ or $Z_F$ seems to slightly change the magnitude of the spin
polarization. However, in metal the interfacial scattering potential
affects the spin polarization a great deal. For a particular value
of $Z$, the increase in $Z_F$ can cause the spin polarization in
metal to change sign.

\section{Conclusions}
According to the results from our simple model, one can directly use
in-plane tunneling conductance spectrum to measure the Rashba energy
of a system with the Rashba spin-orbit coupling. The energy
difference between the onset and the discontinuity in slope of the
conductance spectrum is equal to the Rashba energy. Both features
are found to be robust against variation in the quality of the
junction.

Experimentally, to be able to measure the Rashba energy, the
required energy resolution is at least of the order of the Rashba
energy itself and the temperature is low enough in order that both
features are visible. The Rashba energies in semiconductor-based
heterostructures such as InAs, InGaAs and GaN, are of order 1
meV~\cite{sasa,grundler,matsu,koga2,fuji,ikai}, whereas those of
surface alloys like Li/W(110), Pb/Ag(111), and Bi/Ag(111) can be as
large as 200 meV~\cite{ast,roten,pacile,ast2}. These conditions can
be readily met in modern tunneling measurements~\cite{wolf}.

We also found that as the current is driven through the system, an
imbalance of spin in both sides occurs. The spin polarization of the
conductance in the metal is found to depend strongly on both types
of the interfacial scattering and can disappear when the barrier is
in the tunneling regimes. On the contrary, in the Rashba system the
spin polarization is always present and only slightly affected by
interfacial scattering. This finding suggests that the spin
imbalance caused by current flow in the system with the Rashba
spin-orbit coupling is robust against variation in the quality of
the junction as well.
\begin{acknowledgements}
We thank Dr. Michael F. Smith for critical reading of the
manuscript. Also, we would like to acknowledge financial support
from Cooperative Research Network (Physics). P.P. thanks Thai
Research Fund and Commission on Higher Education, Thailand (grant
no. RMU488012 and CHE-RES-RG "Theoretical Physics") for financial
support. M.B. was supported by the Research Corporation, CIFAR and
NSERC.
\end{acknowledgements}

\end{document}